\documentclass[9pt, conference]{IEEEtran}
\IEEEoverridecommandlockouts
\usepackage{cite}
\usepackage{amsmath,amssymb,amsfonts}
\usepackage{algorithmic}
\usepackage{graphicx}
\usepackage{textcomp}
\usepackage{xcolor}
\usepackage{blindtext}
\usepackage{rotating}
\usepackage{float}

\usepackage{hyperref}
\usepackage{pgfplots}
\usepackage{xcolor}
\usepackage{booktabs} 
\usepackage{bbm}
\usepackage{bm}
\usepackage{multirow}
\usepackage{siunitx}
\usepackage{pifont}            
\newcommand{\cmark}{\ding{51}} 
\newcommand{\xmark}{\ding{55}}
\usepackage[utf8]{inputenc}
\usepackage{amssymb}

\def\BibTeX{{\rm B\kern-.05em{\sc i\kern-.025em b}\kern-.08em
    T\kern-.1667em\lower.7ex\hbox{E}\kern-.125emX}}
\begin{document}

\title{HighRateMOS: Sampling-Rate Aware Modeling for Speech Quality Assessment
\thanks{$^\dagger$\tiny Equal Contribution to this work.}}

% \title{\color{blue}EMO-Debias: Pioneering Benchmark for Multi-Label Speech Emotion Recognition Gender Debiasing Beyond Single-Label}
%%%%%%%%%%%%%%%%%%%%%%%%%%%%%%%%%%%%%%%%%%%%%%%
% \author{\IEEEauthorblockN{Wenze Ren}
% \IEEEauthorblockA{
% \textit{National Taiwan University}\\
% Taipei, Taiwan \\
% email address or ORCID}
% \and
% \IEEEauthorblockN{Yi-Cheng Lin}
% \IEEEauthorblockA{
% \textit{National Taiwan University}\\
% Taipei, Taiwan \\
% f12942075@ntu.edu.tw}
% \and
% \IEEEauthorblockN{Wen-chin Huang}
% \IEEEauthorblockA{
% \textit{Nagoya University} \\
% Nagoya, Japan \\
% wen.chinhuang@g.sp.m.is.nagoya-u.ac.jp}
% \and
% \IEEEauthorblockN{Ryandhimas E. Zezario}
% \IEEEauthorblockA{
% \textit{Academia Sinica}\\
% Taipei, Taiwan \\
% ryandhimas@citi.sinica.edu.tw}
% \and
% \IEEEauthorblockN{Szu-Wei Fu}
% \IEEEauthorblockA{
% \textit{NVIDIA}\\
% Taipei, Taiwan \\
% szuweif@nvidia.com}
% \and
% \IEEEauthorblockN{Sung-Feng Huang}
% \IEEEauthorblockA{
% \textit{NVIDIA}\\
% Taipei, Taiwan \\
% sungfengh@nvidia.com}
% \and
% \IEEEauthorblockN{Erica Cooper}
% \IEEEauthorblockA{
% \textit{name of organization (of Aff.)}\\
% City, Country \\
% email address or ORCID
% }
% \and
% \IEEEauthorblockN{Yu Tsao}
% \IEEEauthorblockA{
% \textit{Academia Sinica}\\
% Taipei, Taiwan \\
% yu.tsao@citi.sinica.edu.tw
% }
% \and
% \IEEEauthorblockN{Hung-yi Lee}
% \IEEEauthorblockA{
% \textit{National Taiwan University}\\
% Taipei, Taiwan \\
% hunuyilee@ntu.edu.tw
% }
% }
% \maketitle
%%%%%%%%%%%%%%%%%%%%%%%%%%%%%%%%%%%%%%%%%%%%%%
\author{
Wenze Ren$^{1\dagger}$, Yi-Cheng Lin$^{1\dagger}$, Wen-Chin Huang$^2$, Ryandhimas E. Zezario$^{3}$, Szu-Wei Fu$^{4}$, Sung-Feng Huang$^{4}$, \\ Erica Cooper$^5$, Haibin Wu$^6$, Hung-Yu Wei$^{1}$, Hsin-Min Wang$^3$, Hung-yi Lee$^{1}$, Yu Tsao$^3$ 
\\
\textit{$^1$National Taiwan University}, \textit{$^2$Nagoya University}, \textit{$^3$Academia Sinica}, \textit{$^4$NVIDIA}\\
\textit{$^5$National Institute of Information and Communications Technology}, \textit{$^6$Independent Researcher}\\

\texttt{\{r11942166, f12942075, hungyilee\}@ntu.edu.tw, yu.tsao@citi.sinica.edu.tw}\\
% $^\dagger$Equal Contribution to this work
}
\maketitle
%%%%%%%%%%%%%%%%%%%%%%%%%%%%%%%%%%%%%%%%%%%%%%
\begin{abstract}
Modern speech quality prediction models are trained on audio data resampled to a specific sampling rate. When faced with higher-rate audio at test time, these models can produce biased scores. We introduce HighRateMOS, the first non-intrusive mean opinion score (MOS) model that explicitly considers sampling rate. HighRateMOS ensembles three model variants that exploit the following information: (i) a learnable embedding of speech sampling rate, (ii) Wav2vec 2.0 self-supervised embeddings, (iii) multi-scale CNN spectral features, and (iv) MFCC features. In AudioMOS 2025 Track3, HighRateMOS ranked first in five out of eight metrics. Our experiments confirm that modeling the sampling rate directly leads to more robust and sampling-rate-agnostic speech quality predictions.
\end{abstract}


\section{Introduction}
Speech quality is critical for modern communication and multimedia systems, as degradation of audio quality leads to loss of intelligibility and poor user experience. To improve audio quality, recent Text-to-Speech (TTS) systems produce speech at sampling rates higher than 16 kHz. For example, Parallel WaveGAN \cite{parallel_wavegan} generates 24 kHz audio in a Transformer-based TTS framework, while HiFi-GAN \cite{hifigan} synthesizes 22.05 kHz speech. As TTS moves to higher sampling rates, evaluations must handle diverse sampling rates to account for differences in frequency range and listening performance.

Most non-intrusive neural predictors of speech quality are trained on data with a single sampling rate. For example, Quality-Net \cite{quality_net}, SSL-MOS \cite{ssl_mos}, MOSNet \cite{mos_net}, and MOSA-Net \cite{mosanet} were trained on speech uniformly downsampled to 16 kHz. This leaves the question of how well these models perform on audio recorded or played back at other sampling rates.

\begin{figure}
    \centering
    \includegraphics[width=0.9\linewidth]{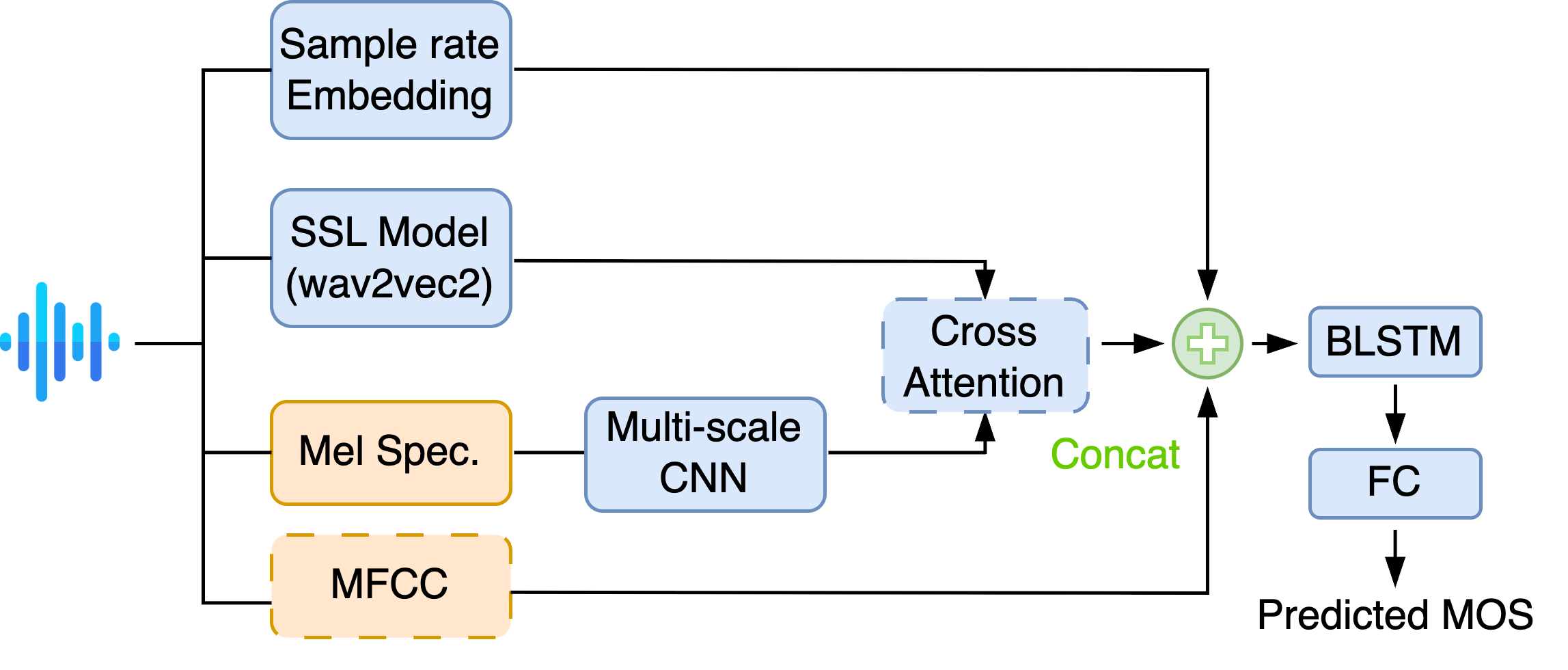}
    \caption{Proposed model architecture. Three models are trained. Model 1 does not include the cross-attention module and MFCC, Model 2 adds the cross-attention module to Model 1, and Model 3 additionally incorporates MFCC training with all blocks.}
    \vspace{-10pt}
    \label{fig:model-architecture}
\end{figure}

In this work, we propose \textbf{HighRateMOS}, an ensemble of three models to predict speech quality at different sampling rates. Each model takes sampling rate embeddings, self-supervised learning (SSL) model representations, and Mel-spectrograms as input features. The complete architecture is depicted in Figure~\ref{fig:model-architecture}. 
HighRateMOS achieved first place in five of the eight evaluation metrics in Track 3 of the AudioMOS Challenge 2025. Compared to the baseline system, the system-level Spearman Rank Correlation Coefficient (SRCC) is improved by 27.5\% and Kendall’s $\tau$ (KTAU) is improved by 53.9\%.

% Incorporate related works here
% Novelty 
% 1. Not downsample, use sampling rate embedding
% 2. Multi-scale CNN
% 3. Comprehensive comparison

\section{Methodology}

\subsection{Feature Extraction} 
We encode the sampling rate (16 kHz, 24 kHz, or 48 kHz) of each utterance into a learned embedding vector, which allows the model to adjust to the sampling rate of the input signal. To capture a broad range of acoustic information, we use SSL embeddings along with Mel-spectrograms and MFCCs. Pretrained SSL models such as wav2vec 2.0 \cite{wav2vec2}, HuBERT \cite{hubert}, and WavLM \cite{wavlm} are trained on raw waveforms sampled at 16 kHz, and their convolutional front ends assume this sampling rate. In previous studies, audio recorded at 24 kHz or 48 kHz is first converted to 16 kHz using downsampling. We posit that an SSL encoder pretrained at 16 kHz can treat audio at a higher sampling rate as the same content played at a slower speed. For example, a 48 kHz waveform resembles 16 kHz speech stretched three times, with the same temporal pattern as the encoder. Because SSL models are trained on large and diverse corpora covering a wide range of speaking rates and acoustic conditions, they can characterize content even when the speech rate is slower.

% For the Mel-spectrogram, we use a multi-scale CNN to capture both fine-grained and broad spectral–temporal patterns that affect perceived speech quality. Convolutional filters at different time–frequency scales enable the model to detect narrowband artifacts (e.g., high-frequency noise) alongside wider distortions (e.g., temporal smearing) within the same architecture.
For the Mel-spectrogram, we use a multi-scale CNN to capture both fine-grained and broad spectral–temporal patterns that affect perceived speech quality. Convolutional filters at different time–frequency scales enable the model to detect narrowband artifacts alongside wider distortions within the same architecture.

\subsection{Feature Aggregation}
In this challenge, we trained three models. As shown in Figure~\ref{fig:model-architecture}, Model 1 takes an intuitive approach by concatenating the frame-level outputs of the SSL encoder, the sampling rate vector, and the multi-scale Mel-spectrograms along the channel dimension to obtain a unified representation; Model 2 introduces a cross-attention mechanism on top of Model 1, which selectively focuses on relevant acoustic patterns from the self-supervised features and the spectral representations; Model 3 integrates all components. After combining multiple features, temporal aggregation is performed via BLSTM, and the final MOS prediction is obtained by a fully connected layer. This design ensures that each model variant can capture different aspects of speech quality.

\subsection{Loss Function}
We divide the eight losses into four categories: point-wise regression loss, ranking-based loss, correlation-based loss, and hybrid loss.
\subsubsection{Point-wise regression loss} We explore the Mean Squared Error (MSE) \cite{quality_net} and Mean Absolute Error (MAE) \cite{9746395} for minimizing the average squared/absolute difference.

\subsubsection{Ranking-based loss}
Ranking-based loss functions encourage the model to preserve the relative ordering of samples by MOS. We employ the contrastive loss \cite{cdpam}, which encourages the model not only to predict correct absolute MOS values but also to match the relative difference between any two utterances. 
% Given a batch of \(n\) samples with ground-truth scores $
% \mathbf{y} = [y_1, y_2, \dots, y_n] $
% and predicted scores $
% \hat{\mathbf{y}} = [\hat y_1, \hat y_2, \dots, \hat y_n],
% $, 
% we compute the pairwise difference matrices
% \[
% \Delta y_{ij} = y_i - y_j,
% \quad
% \Delta \hat y_{ij} = \hat y_i - \hat y_j.
% \]
% The contrastive loss is defined as
% \[
% \mathcal{L}_{\mathrm{cont}}
% = \frac{1}{2\,n^2}
% \sum_{i=1}^{n}
% \sum_{j=1}^{n}
% \max\bigl(0,\;|\Delta \hat y_{ij} - \Delta y_{ij}| - m\bigr),
% \]
% where \(m \ge 0\) is a margin hyperparameter that controls the tolerance for small differences. 
We also employ the Relative Ranking Loss (RelRank), which is the triplet loss among extreme values in a batch \cite{relrank}.
\subsubsection{Correlation-based loss}
Correlation-based loss functions directly optimize the statistical agreement between predictions and human ratings. We explore Linear Correlation Coefficient (LCC) loss (defined as one minus the Pearson linear correlation between predicted and reference MOS) and Concordance Correlation Coefficient (CCC) loss (defined as one minus the CCC), which jointly maximizes both precision and accuracy of the predictions.
\subsubsection{Hybrid loss}
Hybrid losses aim to leverage the complementary strengths of different objectives. We also incorporate widely used losses such as Dual Criterion Quality (DCQ) loss \cite{dcq}, which jointly enforces both precise numerical estimation and correct ordinal ranking, and UTMOS loss \cite{utmos}, which is the sum of clipped MSE loss and $0.5\times$ contrastive loss. We use the same clipping parameter of 0.25 as the original paper.

% Briefly describe the model architecture

\section{Experimental setup}
\subsection{Dataset}
We use the AudioMOS 2025 Track 3 dataset, which contains 1,200 speech samples produced by TTS and neural vocoder systems, each rated for naturalness on a 1-5 Likert scale by 10 annotators. The dataset covers three sampling rates: 16 kHz, 24 kHz, and 48 kHz. The training and development sets share the same 400 utterances but use different listening protocols. In the training set, speech samples at each sampling rate were evaluated in separate sessions, so listeners only compared stimuli with the same sampling rate. Unlike the training set, in the development set, speech samples of the three sampling rates were interleaved in a listening session. The evaluation set follows the same mixed-rate protocol as the development set, requiring listeners to rate stimuli of 16 kHz, 24 kHz, and 48 kHz in one session.

We report two sets of results: development performance during training and final evaluation performance. 
During the training phase, since labels for the development set are not available, we can only obtain the performance of the model on it by uploading predictions to the challenge platform. 
We split the training set into five folds and perform five-fold cross-validation, training on four folds and validating on the held-out fold. 
We then aggregate the predictions from the five held-out folds to evaluate the development performance.
In the evaluation phase, we follow the standard protocol of training on the full training set, selecting the best model on the development set (the labels are available in the evaluation phase), and reporting results on the evaluation set.

\subsection{Models}
We use the SHEET\footnote{\tiny https://github.com/unilight/sheet} \cite{mosbench} toolkit and SSL feature extractors from the S3PRL\footnote{\tiny https://github.com/s3prl/s3prl} \cite{yang21c_interspeech} toolkit to implement our models. 
The SSL models explored include wav2vec 2.0 Base (W2V2), Large (W2V2 L) \cite{wav2vec2}; WavLM Base+ (WavLM B), Large (WavLM L) \cite{wavlm}; Data2vec Large (D2V) \cite{data2vec}, and XLSR-1B (XLSR) \cite{xlsr}. 
We use AdamW \cite{adamw} optimizer with a learning rate of 1e-3, and train the model until reaching system-level SRCC convergence (stops increasing for 2000 steps).
The batch size is 8. 

\subsection{Evaluation Metrics}
We employ a set of metrics at the utterance and system levels, including MSE for measuring prediction accuracy, LCC for evaluating linear relationships, SRCC for capturing monotonic associations, and KTAU for evaluating the ordinal agreement between predicted scores and ground truth scores \cite{voicemos2024}.
\vspace{-5pt}
\begin{table}[ht]
\centering
\caption{Official evaluation results of all submitted systems for AudioMOS Track 3. \textbf{Bold}: the best results.}
\vspace{-6pt}
\setlength{\tabcolsep}{4pt}
\begin{tabular}{l|cccc|cccc}
\toprule
\textbf{} & \multicolumn{4}{c|}{\textbf{Utterance-level}} & \multicolumn{4}{c}{\textbf{System-level}} \\
\textbf{System} & \textbf{MSE} & \textbf{LCC} & \textbf{SRCC} & \textbf{KTAU} & \textbf{MSE} & \textbf{LCC} & \textbf{SRCC} & \textbf{KTAU} \\
\midrule
B03 & 0.273 & 0.821 & 0.695 & 0.508 & 0.119 & 0.941 & 0.749 & 0.547 \\
T01 & 0.303 & 0.804 & 0.643 & 0.459 & 0.104 & 0.957 & 0.866 & 0.705 \\
T08 & 0.277 & 0.811 & 0.716 & 0.529 & \textbf{0.056} & 0.978 & 0.913 & 0.758 \\
T11 & 0.282 & 0.813 & 0.714 & 0.536 & 0.085 & 0.968 & 0.917 & 0.789 \\
T13 & 0.298 & 0.796 & 0.671 & 0.487 & 0.090 & 0.972 & 0.926 & 0.779 \\
T16 & 0.287 & 0.830 & 0.723 & \textbf{0.589} & 0.071 & 0.952 & 0.891 & 0.750 \\
T19 & \textbf{0.238} & 0.846 & 0.694 & 0.513 & 0.080 & 0.955 & 0.914 & 0.758 \\
\midrule
Ours & 0.303 & \textbf{0.847} & \textbf{0.742} & 0.556 & 0.116 & \textbf{0.982} & \textbf{0.955} & \textbf{0.842} \\
\bottomrule
\end{tabular}
\label{tab:challenge_performance}
\vspace{-15pt}
\end{table}
\vspace{-3pt}
\begin{table}[ht]
\centering
\caption{Comparison between losses. \textbf{Bold}: best; \underline{underline}: second‐best.}
\vspace{-6pt}
\setlength{\tabcolsep}{4pt}
\begin{tabular}{@{}l|cccc|cccc@{}}
\toprule
 & \multicolumn{4}{c|}{\textbf{Development}} & \multicolumn{4}{c}{\textbf{Evaluation}} \\
\textbf{Method}      & \textbf{MSE}    & \textbf{LCC}    & \textbf{SRCC}   & \textbf{KTAU}   & \textbf{MSE}    & \textbf{LCC}    & \textbf{SRCC}   & \textbf{KTAU} \\
\midrule
MAE         & 0.095  & 0.942  & 0.768        & 0.611        & 0.130  & 0.958         & 0.926        & 0.768        \\
MSE         & 0.081  & 0.955  & 0.779        & 0.632        & 0.136  & 0.956         & 0.905        & 0.758        \\
RelRank     & \underline{0.049} & 0.954  & 0.824        & 0.653        & 0.088  & 0.958         & 0.929        & 0.800        \\
LCC         & 0.545  & 0.938  & \textbf{0.887} & \textbf{0.811} & 0.424  & 0.937         & \textbf{0.962} & \textbf{0.863} \\
CCC         & 0.069  & \underline{0.969} & 0.836        & 0.674        & \textbf{0.063} & \underline{0.973} & \underline{0.956} & \underline{0.832} \\
DCQ         & \textbf{0.041} & 0.967  & 0.863        & 0.663        & \underline{0.067} & \underline{0.973} & 0.946        & 0.811        \\
UTMOS       & 0.092  & 0.947  & 0.773        & 0.621        & 0.117  & 0.962         & 0.920        & 0.779        \\
\midrule
contrastive & \underline{0.049} & \textbf{0.972} & \underline{0.880} & \underline{0.716} & 0.124  & \textbf{0.982} & 0.949        & 0.821        \\
\bottomrule
\end{tabular}
\label{tab:loss_comparison}
\vspace{-10pt}
\end{table}

\vspace{-4pt}
\begin{table}[htbp!]
\centering
\caption{System‐level performance of different SSL upstreams on development and evaluation sets. \textbf{Bold}: best; \underline{underline}: second‐best.}
\vspace{-6pt}
\setlength{\tabcolsep}{4pt}
\begin{tabular}{@{}l|cccc|cccc@{}}
\toprule
 & \multicolumn{4}{c|}{\textbf{Development}} & \multicolumn{4}{c}{\textbf{Evaluation}} \\
\textbf{Upstream} & \textbf{MSE} & \textbf{LCC} & \textbf{SRCC} & \textbf{KTAU} & \textbf{MSE} & \textbf{LCC} & \textbf{SRCC} & \textbf{KTAU} \\
\midrule
W2V2~L   & 0.055 & 0.969 & 0.832 & 0.663 & \underline{0.047} & 0.984 & \textbf{0.980} & \textbf{0.905} \\
D2V~L    & 0.070 & 0.949 & 0.814 & 0.684 & 0.239 & \underline{0.986} & \underline{0.976} & \underline{0.884} \\
XLSR     & \textbf{0.020} & \textbf{0.980} & \underline{0.872} & \textbf{0.726} & 0.199 & \textbf{0.987} & \underline{0.976} & 0.874 \\
WavLM~L  & 0.036 & 0.964 & 0.812 & 0.663 & 0.100 & 0.975 & 0.956 & 0.832 \\
WavLM~B  & \underline{0.033} & 0.968 & 0.862 & \underline{0.716} & \textbf{0.040} & 0.978 & 0.941 & 0.800 \\
\midrule
W2V2 & 0.049 & \underline{0.972} & \textbf{0.880} & \underline{0.716} & 0.124 & 0.982 & 0.949 & 0.821 \\
\bottomrule
\end{tabular}
\label{tab:ssl_upstream}
\end{table}
\vspace{-3pt}

%%%%%%%%%%%%%%%%%%%%%%%%%%%%%%%%%%%%%%%%%%%%%%%%%%%%%%%%%%%
\begin{table*}[!htbp]
  \centering
  \caption{
    Ablation study on model components on the evaluation set. \textbf{Bold}: the best results, \textbf{\cmark} and \textbf{\xmark} : retained and removed components.}
    \vspace{-6pt}
  \setlength{\tabcolsep}{6pt}
  \footnotesize
  \begin{tabular}{ccccccc|cccc|cccc}
    \toprule
    \textbf{} & \textbf{} & \textbf{} &
    \textbf{} & \textbf{} & \textbf{} & \textbf{} &
    \multicolumn{4}{c|}{\textbf{Utterance-level}} &
    \multicolumn{4}{c}{\textbf{System-level}} \\
    \textbf{SSL} & \textbf{SR Emb} & \textbf{Mel} & \textbf{Multi-CNN} & \textbf{MFCC} & \textbf{Cross-Attn} & \textbf{BLSTM} &
    \textbf{MSE} & \textbf{LCC} & \textbf{SRCC} & \textbf{KTAU} &
    \textbf{MSE} & \textbf{LCC} & \textbf{SRCC} & \textbf{KTAU} \\
    \midrule
    \midrule
    \cmark & \cmark & \cmark & \cmark & \cmark & \cmark & \cmark &
    0.537 & 0.761 & 0.666 & 0.477 & 0.247 & 0.982 & 0.956 & 0.842 \\

    \cmark & \cmark & \cmark & \cmark & \xmark & \xmark & \cmark &
    0.360 & 0.821 & 0.720 & 0.533 & 0.187 & 0.975 & 0.919 & 0.779 \\

    \xmark & \cmark & \cmark & \cmark & \cmark & \cmark & \cmark &
    0.632 & 0.749 & 0.648 & 0.458 & 0.420 & 0.967 & 0.892 & 0.726 \\

    \cmark & \xmark & \cmark & \cmark & \cmark & \cmark & \cmark &
    0.450 & 0.806 & 0.717 & 0.524 & 0.223 & \textbf{0.985} & 0.955 & 0.842 \\

    \cmark & \cmark & \xmark & \xmark & \cmark & \cmark & \cmark &
    0.377 & 0.762 & 0.675 & 0.484 & 0.107 & 0.975 & 0.938 & 0.800 \\

    \cmark & \cmark & \cmark & \xmark & \cmark & \cmark & \cmark &
    0.541 & 0.772 & 0.662 & 0.470 & 0.322 & 0.967 & 0.902 & 0.737 \\

    \cmark & \cmark & \cmark & \cmark & \xmark & \cmark & \cmark &
    \textbf{0.276} & \textbf{0.837} & \textbf{0.724} & \textbf{0.540} & \textbf{0.063} & 0.973 & 0.943 & 0.811 \\

     \cmark & \cmark & \cmark & \cmark & \cmark & \xmark & \cmark &
    0.601 & 0.798 & 0.704 & 0.513 & 0.352 & 0.975 & 0.949 & 0.832 \\

    \cmark & \cmark & \cmark & \cmark & \cmark & \cmark & \xmark &
    0.321 & 0.801 & 0.689 & 0.502 & 0.084 & \textbf{0.985} & \textbf{0.981} & \textbf{0.905} \\
    \bottomrule
  \end{tabular}
  \label{tab:ablation_components}
  \vspace{-10pt}
\end{table*}
%%%%%%%%%%%%%%%%%%%%%%%%%%%%%%%%%%%%%%%%%%%%%%%%%%%%%%%%%%%%%
\vspace{-3pt}
\begin{table}[htbp!]
\centering
\caption{Comparison across different ensemble configurations, \textbf{Bold}: the best results.}
\vspace{-6pt}
\setlength{\tabcolsep}{2.5pt}
\begin{tabular}{@{}l|cccc|cccc@{}}
\toprule
 & \multicolumn{4}{c|}{\textbf{Development}} & \multicolumn{4}{c}{\textbf{Evaluation}} \\
\textbf{Ensemble} & \textbf{MSE} & \textbf{LCC} & \textbf{SRCC} & \textbf{KTAU} & \textbf{MSE} & \textbf{LCC} & \textbf{SRCC} & \textbf{KTAU} \\
\midrule
Setting 1   & 0.343 & 0.832 & 0.717 & 0.531 & 0.129 & 0.978 & 0.950 & 0.832 \\
Setting 2    & 0.279 & 0.835 & 0.712 & 0.521 & 0.133 & 0.967 & 0.922 & 0.768 \\
Setting 3     & 0.333 & 0.840 & 0.726 & 0.541 & 0.144 & 0.980 & 0.949 & 0.821 \\
Setting 4     & 0.372 & 0.822 & 0.704 & 0.518 & 0.204 & 0.957 & 0.893 & 0.737 \\
HighRateMOS  & \textbf{0.303} & \textbf{0.847} & \textbf{0.742} & \textbf{0.556} & \textbf{0.116} & \textbf{0.982} & \textbf{0.955} &  \textbf{0.842}\\
\bottomrule
\end{tabular}
\label{tab:ensemble}
\end{table}
\vspace{-3pt}

\section{AudioMOS 2025 challenge results}
Table~\ref{tab:challenge_performance} shows the official evaluation results of our system on the evaluation set compared with the baseline and other systems. Our system achieved first place in 5 out of 8 metrics and second place in one metric. It is worth noting that our system showed a substantial performance advantage over all other systems in the system-level SRCC and KTAU metrics. 

\section{Experimental Results}
\subsection{Preliminary Study on Losses}

We conducted a preliminary study on SSL-MOS using wav2vec 2.0 to probe the effect of different losses before selecting the loss in our final system. 
Table \ref{tab:loss_comparison} shows that different loss functions yield varied performance trade-offs across metrics. On the development set,  \textbf{contrastive loss} delivers the best LCC, and the second-best SRCC, KTAU, and MSE. 
Given that the primary challenge objective is to maximize system-level SRCC, the \textbf{contrastive loss} emerges as the most suitable choice. 
By comparison, training with an LCC loss yields the highest SRCC but suffers from a very large MSE, indicating poor absolute calibration. 

When development labels are available, LCC loss achieves the highest system SRCC and KTAU on the evaluation set, making it the best choice for pure ranking. 
Contrastive loss remains a robust middle ground, achieving the highest LCC (0.982) and competitive SRCC (0.949), and thus is a strong alternative when both ranking and calibration matter.

\subsection{Preliminary Study on SSL Upstream}
To identify the best SSL encoder for robust MOS prediction across varied sampling rates, we compared several SSL encoders using contrastive loss. In Table \ref{tab:ssl_upstream}, the W2V2 achieves an exceptional development‐set ranking capability, with the highest SRCC and the second‐highest LCC. This led us to select W2V2 as our upstream model.
However, our contrastive‐trained W2V2 on the evaluation set does not outperform all other models. W2V2 L achieves the highest SRCC, while XLSR attains the highest LCC. This may reflect the distribution shift between training and development/evaluation data.

\subsection{Ablation Study}
% Performance of architecture 
% 1. without sampling rate embedding
% 2. Multi-CNN * 3 / Multi-CNN / single CNN
% 3. +/- MFCC

% To ensure the reliability and reproducibility of the results, see 
We retrained the proposed model and conducted a comprehensive ablation by removing each of the seven core components (SSL features, sampling rate embedding, Mel spectrogram features, multi-scale CNN, MFCC features, cross-attention, and BLSTM) one by one.
Table \ref{tab:ablation_components} shows the results on the evaluation set. SSL features were proven to be the most critical component; removing them drops the system-level SRCC significantly from 0.956 to 0.892, indicating that the pretrained SSL representations provide core acoustic information for cross-sampling rate speech quality prediction.
The joint removal of Mel spectrum maps and multi-scale CNNs also led to a significant decrease in system-level SRCC, proving that spectral features are very important for predicting the quality of audio with different sampling rates. In contrast, the removal of MFCC features, cross-attention mechanisms, and sampling rate embeddings had a relatively small impact on performance.

Removing BLSTM actually improved system performance, a counterintuitive result that suggests that simple FC layers may be more effective than complex sequence modeling in speech quality assessment tasks. We believe this happens because the BLSTM tended to overfit on our relatively small training set, whereas the FC layers remained more robust.

\subsection{Ensemble Strategy}
In Table \ref{tab:ensemble}, we explored four ensemble configurations combining our three models. The “5-fold” refers to the evaluation scores of the five cross-fold validation models trained during the training phase.
\textbf{Setting 1}: Model 1 (5-fold average) + Models 2 and 3 (standard training).
\textbf{Setting 2}: All models use 5-fold averaging.
\textbf{Setting 3}: All models use standard training. 
\textbf{Setting 4}: All models use the best among 5-fold.
\textbf{HighRateMOS}: Model 1 (best among 5-fold) + Models 2 and 3 (standard training). HighRateMOS significantly outperforms other configurations and achieves the highest evaluation performance. Since Model 1 shows significant performance variation across its cross-validation folds, choosing its single best fold captures the configuration most aware of sampling-rate differences. This validates that our hybrid integration method is very effective for speech quality prediction across different sampling rates.

\section{Conclusion}
HighRateMOS tackles the distribution shift that plagues non-intrusive speech-quality prediction when signals arrive at different sampling rates by unifying three key ideas: explicit sampling-rate embeddings, multi-scale feature fusion, and ensemble averaging. It encodes 16 kHz, 24 kHz, and 48 kHz as learnable vectors, combines wav2vec 2.0’s self-supervised acoustic representations with Mel and MFCC spectral details through cross-attention and BLSTM aggregation, and then averages the outputs of complementary model variants. This design preserves the synergy between high-level acoustic and fine-grained spectral cues while sharply reducing cross-bandwidth prediction drift. On AudioMOS 2025 Track 3, HighRateMOS delivers a system-level SRCC of 0.955 and KTAU of 0.842, significantly exceeding the baseline and other systems. Although its performance is still limited by the size and linguistic diversity of the training data, this work demonstrates that explicit sampling-rate modeling is a practical approach to bandwidth-agnostic, robust speech-quality assessment, setting the stage for future advances such as continuous-rate encoding and multilingual expansion.

\bibliographystyle{IEEEtran}
\bibliography{mybib}

\end{document}